\documentclass[twocolumn,showpacs,preprintnumbers,amsmath,amssymb]{revtex4}
\usepackage{graphicx}% Include figure files
\usepackage{dcolumn}% Align table columns on decimal point
\usepackage{bm}% bold math
\usepackage{ams}% ams math symbols

\begin{document}
\newcommand{\heyon}{$^4$He }
\newcommand{\hesan}{$^3$He }
\newcommand{\kake}{$\times$}

%\title{Creeps in Dissipation and Non-Classical Rotational Inertia in %Supersolid \heyon}% Force line breaks
%\title{Dissipation Relaxation Dynamics of Transiently Stressed Solid \heyon}
%\title{Dissipation Relaxation Dynamics in Oscillation Solid \heyon at Low Temperatures}
%\title{Dissipation Relaxation Dynamics of Torsionally Driven Solid \heyon}
%
\title{New Dissipation Relaxation Phenomenon in Oscillating Solid \heyon}
\author{Y. Aoki}
\author{M.C. Keiderling}
\author{H. Kojima}
\affiliation{Serin Physics Laboratory, Rutgers University, Piscataway,
NJ 08854 USA}

\date{\today}% It is always \today, today,
             %  but any date may be explicitly specified

\begin{abstract}
We describe the first observations on the time-dependent dissipation when the drive level of a torsional oscillator containing solid \heyon is abruptly changed.  The relaxation of dissipation in solid \heyon shows rich dynamical behavior including exponential and logarithmic time-dependent decays, hysteresis, and memory effects.
\end{abstract}

% PACS, the Physics and Astronomy
                             % Classification Scheme.
%\keywords{Suggested keywords}%Use showkeys class option if keyword
                              %display desired
\maketitle
%\subsection{introduction/motivation}
The recent discovery \cite{Kim04a,Kim04b} and its confirmations \cite{Rittner06,Kondo07,Penzev07,Aoki07} of significant low temperature reductions in the rotational moment of inertia of solid \heyon below 300 mK have generated enormous excitement as possible evidence for the long-sought supersolid behavior \cite{Prokofev05}.  The physical origin of the apparent non-classical rotational inertia (NCRI) extracted from the temperature dependent shift in the torsional oscillator frequency, however, remains unsettled \cite{Balibar08}.  The majority of the torsional oscillator experiments showing NCRI has been devoted to measuring the oscillator frequency under a variety of conditions in such as crystal quality \cite{Rittner06}, sample size and geometry, and \hesan impurity concentration \cite{Kim08}.  Relatively little experimental information is available on the dissipation accompanying the torsional oscillation.
%
%Relatively little attention has been paid to the temporal dependence of the NCRI phenomenon of both the frequency shift and the dissipation in the torsional oscillator.
%
If the observed reductions in rotational inertia are related, for example, to vortex liquid/solid phenomenon \cite{Anderson07,Huse07}, glassy behavior \cite{Boninsegni06b,Nussinov07}, or defect motion \cite{Day07} in solid $^4$He, then dynamical time-dependent effects might occur in the dissipation in the oscillating solid \heyon samples.  We carried out systematic observations on the time-dependent dissipation accompanying sudden changes in the torsional oscillator displacement amplitude.  We have discovered that the dissipation in solid \heyon shows rich dynamical behavior including exponential and logarithmic time-dependent relaxation, hysteresis and memory effects.

%\subsection{decay in dissipation and frequency shift}
The torsional oscillator utilized was the compound oscillator identical to that described in \cite{Aoki07}.  Briefly, our oscillator contained an upper "dummy bob" hung by an upper rod and connected to a lower rod attached to the sample chamber containing a cylindrical (10 mm diameter and 8mm height) solid \heyon.   The oscillator resonated at "in-phase" and "out-phase" modes at frequencies, $f_1$ = 496 Hz and $f_2$ = 1173 Hz, respectively, whose amplitudes and phases were continuously tracked. The temperature was measured by a calibrated \hesan melting pressure thermometer mounted onto the vibration isolation copper block to which our oscillator was attached.  All of our solid samples were grown from commercially available \heyon gas with nominal 0.3 ppm \hesan impurity by the usual blocked capillary method.

\begin{figure}[htbp]
\includegraphics[width=3in]{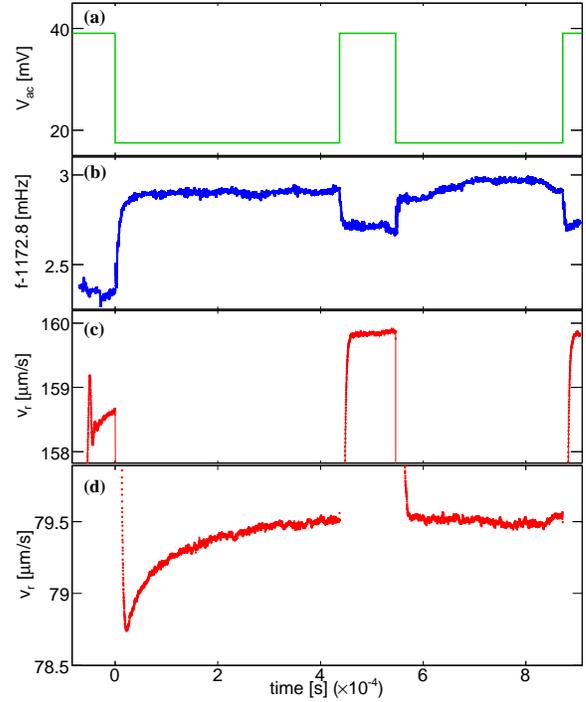}
%example_decay_v2.eps
\caption{\label{example_decay_v2} Time dependence of, ac drive amplitude (a), frequency shift (b), and rim velocity ((c) and (d)) at $T$ = 9.95 mK.  The initial $v_{r}$ ($\sim$160 $\mu$m/s) corresponds to displacement amplitude of 22 nm. The drive (torque) level is $\propto V_{dc}V_{ac}$, where the dc bias voltage $V_{dc}$ is kept constant.  The ring down portion of $v_{r}$ is not shown in (d).  The drive level was increased back to the original level at $t$ = 5$\times$10$^4$ s.  The frequency shift does not return to the initial value in accordance with our observation of hysteresis \cite{Aoki07}.  After the increase in drive, there is no decay similar to that after the first decrease in drive.  The ring down time of the empty oscillator itself ranges from 120 to 200 s.} \end{figure}
\begin{figure}
\includegraphics[width=3in]{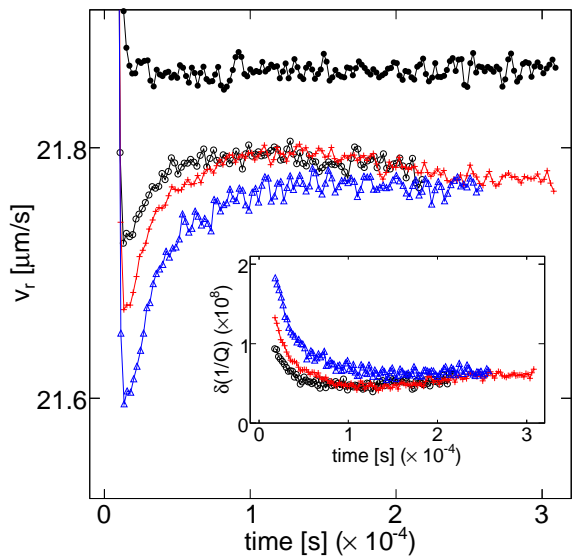}
%amplitude_dissipation.eps
\caption{(color online)Time dependence of sample rim velocity for different initial drive levels at 29.45 mK.  The initial drive level was set such that v$_r$ were 104 (circles), 143 (pluses) and 180 (triangles) $\mu$m/s and the same final v$_r$ $\approx$ 22 $\mu$m/s.  When the cell was filled with superfluid, there was only the oscillator ring down with no anomalous undershoot response (dots).  The inset shows the dissipation due to the presence of solid \heyon sample derived (see text)from the data for each initial v$_r$ in the main panel.}\label{amplitude_dissipation}
\end{figure}

The procedure for measuring the time dependent behavior of our torsional oscillator is illustrated in Fig. \ref{example_decay_v2}.  The $f_2$ mode resonance was continuously tracked by a lock-in amplifier maintaining a constant phase between the drive and the response.  The shifts in resonance frequency and the response amplitude were monitored continuously throughout the procedure.  The response amplitude was converted\cite{Aoki07} to sample rim velocity amplitude, v$_r$.  The cell temperature was initially raised to 90 mK, where the NCRI was found to be reversible as a function of drivel level \cite{Aoki07}.  The oscillator was then driven at a given level and allowed to achieve a steady state.  The temperature was next lowered within about 10$^3$ s to 9.95 mK at the time defined as $t = - t_0 = -5\times 10^3$ s and subsequently regulated within 10 $\mu$K.  The drive level was reduced to half of the initial value at $t = 0$ (see panel (a)).  Following the reduction in drive, the rim velocity (panel (d)) unexpectedly "undershot" and eventually recovered in a complex relaxation process.  The usual ring down of the oscillator resonance transpired during the time interval \textit{prior} to the recovery from the undershooting.  When the drive level was restored to the original value at $t=4.4 \times 10^{4}$ s, v$_r$ increased more than twice but the relaxation behavior was absent.  When the process was repeated to decrease the drive level by half at $t = 5.4 \times 10^{4}$ s, the initial undershoot behavior was now absent but "remembered" its previous state at the same drive level.  Thus, the unexpected relaxation phenomenon occurs only in the \textit{first} decrease from one high drive level to a lower one.  Majority of the relaxation phenomena described here was studied with the $f_2$ mode.  The $f_1$ mode was studied at 9.95 and 19.65 mK and qualitatively similar relaxation phenomena were observed.  When the solid \heyon was replaced with superfluid and the similar procedure was followed, the undershoot and subsequent recovery were both absent at all temperatures studied (see Fig. \ref{amplitude_dissipation}).  This provided clear evidence that the observed unusual time dependence in the oscillator response originates in the solid \heyon sample.

During the observed decay process, the oscillator frequency also showed time dependence (see Fig. \ref{example_decay_v2}(b)).  When $v_{r}$ was initially decreased, the oscillator frequency (therefore, the NCRI fraction) increased.  This is the "critical velocity" effect as seen previously \cite{Kim04b,Aoki07}.  There was an associated relaxation in frequency shift which appeared with similar dynamical behavior as v$_r$.  Owing to the much smaller effect on the frequency relaxation, we focused on the much larger effects of dissipation dynamics.  Observations on the frequency relaxation resulting from temperature changes were reported recently \cite{Clark07d}.

\begin{figure}[htbp]
\includegraphics[width=3in]{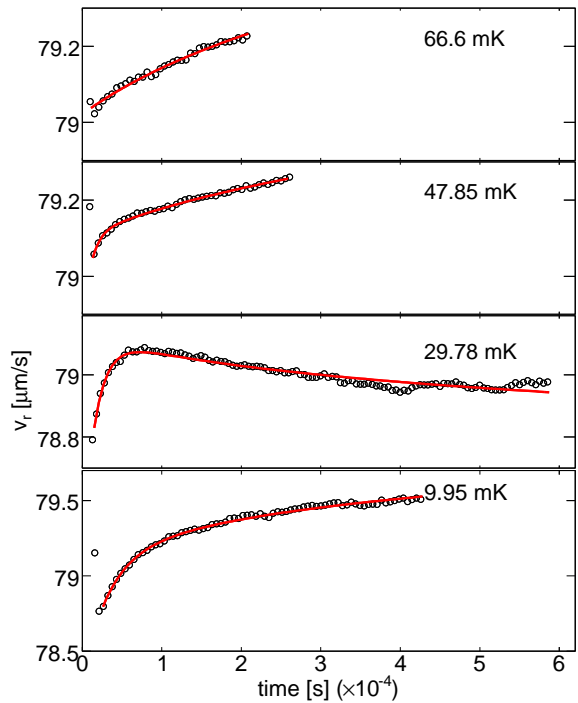}
%decayTdependence.eps
\caption{\label{decayTdependence}Time dependence of the sample rim velocity (v$_r$) at different temperatures.  The same protocol as in Fig. \ref{example_decay_v2} was followed.  The temperatures were 9.95, 29.78, 47.85 and 66.60 mK.  The lines are fits to Eq. (\ref{decay}) discussed in the text.}
\end{figure}

To see how the amplitude relaxation phenomena as shown in Fig. \ref{example_decay_v2} depended on temperature, the same protocol was followed except the final temperature.  Some of such measurements are shown in Fig. \ref{decayTdependence}.  Similar undershoot in rim velocity was seen at temperatures below 50 mK just after the decrease in drive level.  Following the relatively rapid recovery, there was a much slower relaxation at  $t \gtrsim$ 10$^4$ s.  The long time tail behavior after the decrease of drive level was complicated.  In the long time limit, the rim velocity \textit{decreased} in the temperature range, 20 mK $>T>$ 30 mK, but it increased outside of this range.
%\subsection{fit}

There have been suggestions that the supersolid state might involve glassy behavior in some way.  The decay of magnetization of spin glass systems\cite{Nordblad86} has been observed under certain conditions to decay with stretched exponential time dependence. Fitting our data to such time dependence was attempted but inadequate.  We find that a sum of purely exponential and logarithmic time dependencies can fit our data satisfactorily:
\begin{equation}\label{decay}
v_{r}=v_{0} + B\ln (t + t_0) - C\exp (-t/\tau)
\end{equation}
where $v_0$, $B$, $C$ and $\tau$ are fitting parameters.  The short and long time recovery behaviors are described by the exponential and the logarithmic time dependence, respectively.  The parameters $v_0$ and $B$ are first set by fitting the data in the long time limit ($t >$ 1$\times$10$^4$ s), where the exponential term is negligible.  The "origin" of time is taken as that when the final temperature is stabilized ($t = - t_0$) to account for the decay during the waiting time interval $t_0$.  Once the logarithmic decay parameters are set, the remainder is fitted with the exponential dependence in Eq. (\ref{decay}).  The fits (lines) with parameters shown in Fig. \ref{decayTdependence} appear adequate.  The exponential decay amplitude $C$ decreases as the temperature is raised and vanishes in the fit at 66.6 mK in Fig. \ref{decayTdependence}.
%Fig. 2 note: indicate fitting parameters, v_0, B, C and \tau  in each panel.
%note: solid pressure=?

To examine the dependence of the relaxation behavior on the initial drive level, observations were made at 29.45 mK in the same protocol as above keeping the same final v$_r$ at $\sim$ 22 $\mu$m/s but varying the initial v$_r$ between 180 and 100 $\mu$m/s.  As shown in Fig. \ref{amplitude_dissipation}, the undershoot magnitude increased as the difference between the initial and final v$_r$ was increased.  A small dependence on the difference was also present (see below) in the long time tail behavior.  When the cell was filled with superfluid $^4$He, there was no relaxation other than the oscillator ring down.  The inset shows the dissipation in the sample (subsequent to the end of the undershoot) computed from the observed amplitude $A$ ($A_0$) in the presence of solid (superfluid) \heyon sample via $\delta(1/Q) \equiv (1/Q) - (1/Q_0)$, where $Q =aA$ and $Q_0=aA_0$ ($a$ is the measured calibration constant).  The inset demonstrates the unexpected, large transient \emph{excess} dissipation observed during the recovery process just after the change in drive level.
%Replace parts of above description of inset to Fig. \ref{amplitude_dissipation} by: (In the previous torsional oscillator measurements, the dissipation of the solid has a maximum around 60 mK, and it decreases almost to 0 below 40 mK. Our observation was done in the low temperature range where the dissipation of the solid is very small. The inset demonstrates that even though the undershoot amplitude in $v_{r}$ is small, the relative change in the dissipation due to the solid is large. The change in the dissipation increases as the initial $v_{r}$ increases, which indicates that the solid has a larger dissipation as $v_{r}$ increases at $t<0$. This $v_{r}$ dependence of the dissipation is opposite of the previous measurements above 40 mK, and we show a possibility that it changes below 30 mK.)
%--The inset, by itself does not demonstrate the dependence of dissipation on rim velocity.  To show this we need to measure dissipation vs. rim velocity.  This is difficult since the dissipation itself is near zero.
%Fig. 3 note: what was the initial velocity for the superfluid case? pressure=?

\begin{figure}%[htbp]
\includegraphics[width=3in]{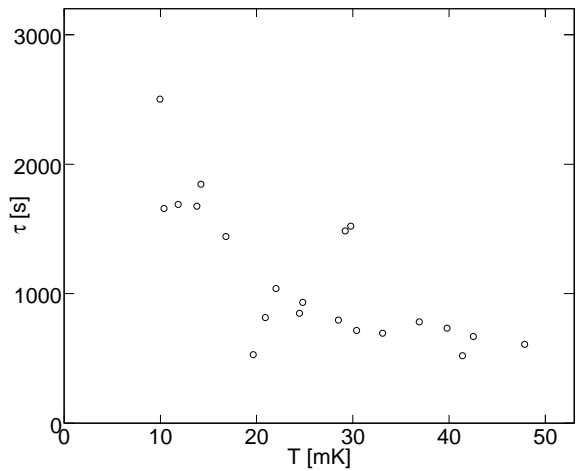}
%exp_relaxation_time.eps
\caption{\label{exp_relaxation_time} Temperature dependence of fitted time constant $\tau$ in Eq. \ref{decay}. This relaxation time does not depend on the drive level.}
\end{figure}

Temperature dependence of the fitted time constant $\tau$ is displayed in Fig. \ref{exp_relaxation_time}, which includes those from the measurements with the same protocol as in Fig. \ref{example_decay_v2} but with data taken only up to about $t$ = 2$\times$10$^4$ s.  As the temperature is increased, $\tau$ decreases monotonically.  The temperature dependence may be represented by $\tau = \tau_0e^{\Delta/(k_BT)}$, $\tau_0$ = 5.4$\times$10$^2$ s and $\Delta/k_B$ = 14 mK.  This characteristic temperature represents the onset of a newly uncovered dissipation dynamics in oscillating solid $^4$He.
% To our knowledge, this is the first observation of a new characteristic energy related to the dissipation in oscillation solid \heyon.
%
%Those time constants fitted from shorter measurement time interval have greater uncertainty.
%The exponential part of the relaxation is hysteretic in a similar manner as the NCRI fraction \cite{Aoki07}.  The coefficient $C$ in Eq. (\ref{decay}) is 1 $\sim$ 2 $\mu$m/s at the lowest temperatures and tends to vanish near 70 mK.

The logarithmic variation in the relaxation in Eq. (\ref{decay}) is a common time dependence to various decay phenomena in superfluids \cite{Reppy70}, superconductors \cite{Anderson64}, frictional processes \cite{Persson95} and glasses \cite{Nordblad86}.  The logarithmic relaxation we observe has complex dependence on temperature and v$_r$.  The sign of the fitted parameter $B$ is negative around 30 mK for the \textit{particular} protocol followed for the data in Fig. \ref{decayTdependence}.  It is positive (negative) when the final v$_r$ is greater (less) than $\approx$ 80 $\mu$m/s, but this rim velocity boundary is dependent on the initial v$_r$ as well as temperature.  The effects of different sample preparation procedures and annealing on the logarithmic decay have not yet been studied. Judging from the hysteresis behavior of NCRI fraction being relatively insensitive to different samples in our cylindrical cell, we expect similar insensitivity in the relaxation phenomena described here.
%
%\begin{figure}[htbp]
%\includegraphics[width=3in]{fig/fig5.eps}
%BvsT.eps
%\caption{\label{BvsT} Temperature dependence of fitted parameter $B$ in Eq. \ref{decay}.  A reentrant behavior is observed over a small temperature range around 25 mK.}
%\end{figure}
%see in C:\...\Torsion Oscillator 1\relaxation effects\relaxation.opj
%
%\begin{figure}[htbp]
%\includegraphics[width=3in]{fig/BvsFinalV.eps} \caption{\label{BvsFinalV} Dependence of fitted time constant $B$ in Eq. \ref{decay} on final rim velocity at $T$ = ? mK.  The initial rim velocity is 237 $\mu$m/s for all data, but the final rim velocity $V_0$ is varied as shown.  New calibration used to evaluate velocities?}
%\end{figure}
%\subsection{aging effect?}
%Is there any evidence for dependence of $\tau$ depends on $t_0$??

The behavior of the parameter $B$ is complicated.  It is known that the particular thermal history (\cite{Clark07d} and our studies) of a solid \heyon sample affects the oscillator \textit{frequency} shift and its relaxation.  For example, the direction of decay depends on the direction of temperature change.  We do not yet have a complete thermal history characterization of the logarithmic decay of the dissipation and it is possible that the observed sign reversal of $B$ is related to some subtle difference in thermal history.
%Clearly, more studies are needed.
%?In the temperature range where the parameter $B$ in the long time limit is found to be negative, is the decay prior to decreasing the drive level also shows negative $B$?  Does the manner of lowering T from high to final temperature affect the sign of $B$?
%
%\subsection{relation to shear modulus and dislocation network idea}
%Recently, Day and Beamish \cite{Day07} found $\sim$ 10 \% increase in the shear modulus of thin solid \heyon slab samples in the same temperature range as the NCRI effects were observed.  The shear modulus showed hysteresis behavior similar what we \cite{Aoki07} had found.  How, if any, their observations are related to NCRI is yet unclear \cite{Clark07d}.  The increase in shear modulus was interpreted \cite{Day07} in terms of

Possible association has been suggested between the observed NCRI phenomena and the dislocation line network depending on the number, $n_3$, of condensed \hesan impurities on the lines \cite{Day07,Kim08}.  We consider the following effect of dislocation on our observed dissipation relaxation phenomena.  Increasing the drive level might partially breakaway \hesan from the dislocation lines.  Then, the distance ($L_i$) between adjacent pinning sites at \hesan locations increases \cite{Iwasa80}.  Just after the drive level is decreased at $t = 0$, $n_3$ is initially too small for the new lower drive level and there is an excess dissipation resulting in the undershooting of the measured amplitude.  As more \hesan recondenses onto the dislocation lines, the dissipation might decrease as seen by the subsequent increase in amplitude.  A difficulty with this scenario is that assuming a uniform dislocation line density $\Lambda$ $\sim$ 10$^5$ cm$^{-2}$\cite{Iwasa80} (average separation between lines $\sim$ 17 $\mu$m) and taking the diffusion coefficient $D$ of \hesan impurities (of concentration $x_3$) in solid \heyon as $Dx_3 \approx 3\times 10^{-11}$ cm$^2$/s at low temperatures \cite{Allen82} give a diffusion time less than 10 s, short compared to $\tau$ in Fig. \ref{exp_relaxation_time}.  The observed $\tau$ might instead be related to the time for \hesan to migrate along dislocation lines towards an equilibrium distribution.

Another possible mechanism for the observed relaxation dynamics is the effect on the number of dislocations induced by the change in oscillation amplitude.  Hiki and Tsuruoka \cite{Hiki83} showed that suddenly imposed changes in temperature $\Delta T$ could exert internal shear stresses and thereby change the number of dislocations.  The measured relaxation in ultrasonic attenuation was interpreted as evidence for quantum mechanical tunneling of dislocations \cite{Hiki83}.  The tunneling \emph{rate} increased as the imposed change in temperature was increased.  If the change in oscillator amplitude in our experiment were equivalent to (but likely to be much less than) $\Delta T$ = (1/10)$T$, the resulting tunneling time constant would be 10$^5$ s, much longer than the observed value.  The quantum mechanical tunneling mechanism also leads to a temperature-independent relaxation time contrary to our finding.  Thus, it is not likely that the relaxation time we observe is related to the dislocation tunneling.  The observed long-term memory effects and logarithmic time dependence are also difficult to explain by the dislocation-related mechanisms.

%We conclude that changes in neither \hesan breakaway nor dislocation density are responsible for the observed relaxation phenomena.
%
%This recondensing process apparently occurs with the observed exponential time dependence.  The temperature dependence of the time constant (see Fig. \ref{exp_relaxation_time}) suggests that the characteristic energy for \hesan condensation is 15 mK.  At low temperatures, where NCRI saturates, the time constant tends to diverge and the dislocations appear totally immobilized in agreement with the suggestion by Kim et al.\cite{Kim08}.
%
%Ultrasonic experiments\cite{Iwasa80,Hiki83} in hcp \heyon with \hesan impurities above 200 mK have shown that the interaction potential between \hesan and dislocations is $\sim$ 0.4 K.  Our observations on the dissipation relaxation dynamics suggest that a different mechanism in $^3$He-dislocation interaction comes into play below $T_{IP}\sim$ 60 mK where impurities dominate the dislocation pinning.
%
%\subsection{physical origin of $\tau$ and vortex liquid}
We discuss next our observations in terms of a vortex liquid model \cite{Anderson07} of supersolid state.  In the model, NCRI is interpreted as the rotational susceptibility of the quantized vortex tangle and the dissipation arises from the interaction between the torsional oscillation and the thermally fluctuating vorticies.  Prior to the first change in the drive level, our procedure of initiating the oscillation at a high temperature is likely to have brought the sample solid \heyon to a quasi-equilibrium array of vortices for the drive level at the temperature of measurement.  The observed logarithmic time dependence probably arises from the thermally activated motion of vortices from one metastable state to another.  This process continues throughout the measurement while the sample is maintained at the constant temperature.  Just after the drive level is decreased, the existing vortices are out of equilibrium with the new drive level.  The observed undershoot may be a transient excess dissipation owing to the "wrong" number of vortices.  The vortices now must adjust to the new drive level by, say, moving out of the sample.  The characteristic time involved in the macroscopic motion of the vortices is then the observed time constant $\tau$.  The motion presumably involves processes occurring both within the sample and at the surface boundaries.  The absence of overshoot and exponential decay in both the oscillator response and NCRI fraction implies that the drive reversal does not replenish vortices into the sample.  This indicates the presence of surface barrier which inhibits creation of vortices in the system.  When the drive level is decreased again, the number of vortices is already near that appropriate for the velocity and there is no undershoot nor accompanying exponential relaxation.  The time constant becomes shorter as the vortices begin to move more freely.  This vortex liquid state appears to occur above about 60 mK.  When the vortex liquid "freezes," the time constant would diverge.  Our sample at the lowest temperature apparently did not attain a totally frozen vortex state.
In conclusion, we studied the dynamics of dissipation imparted by cylindrical solid \heyon samples onto a torsional oscillator by suddenly decreasing the oscillator drive.  Unexpectedly, the rim velocity initially undershoots, recovers from the undershoot exponentially with a characteristic time constant and eventually decays logarithmically in the long time limit.  The time constant tends to diverge as temperature is decreased below 15 mK.

%\subsection{acknowledgment}
We thank J. Graves for developing our compound torsional oscillator.  We thank E. Abrahams, M.H.W. Chan, M. Tachiki and participants at the Pacific Institute of Theoretical Physics workshop in Stillwater for discussions, and M. Paalanen for a correspondence.  This research was supported in part by the NSF through grant DMR-0704120.

\bibliography{supersolid,helium4,material,helium3}% Produces the bibliography via BibTeX.
\end{document}